\documentclass[twoside,a4paper]{article}
\usepackage[T2A]{fontenc}
\usepackage[cp1251]{inputenc}
\usepackage[english]{babel}
\usepackage{graphicx}
\makeatletter
\renewcommand{\@cite}[2]{{#1 \if@tempswa   #2\fi}}
\renewcommand{\@biblabel}[1]{\hfill}
\makeatother

\newcommand{\beq}{\begin{equation}}
\newcommand{\eeq}{\end{equation}}
\newcommand{\bea}{\begin{eqnarray}}
\newcommand{\eea}{\end{eqnarray}}

\begin{document}

\vbox{}

\begin{center}
{\LARGE  \bf Notes on non-thermal X-ray radiation of radio supernova remnant W\,50 and collimated radiation of SS\,433}
\vspace{0.7cm}

{\large  A.~A. Panferov}

{\small IMFIT, Togliatti State University, Russia;
e-mail: panfS@yandex.ru}

{\large  A.~A. Kabanov}

{\small Samara State University, Russia;
e-mail: artkabanov@mail.ru}
\end{center}
\vspace{0.7cm}

\begin{abstract}
Diffuse X-ray emission of the radio nebula W\,50 along the line of the of jets of 
the microquasar SS\,433 has a non-thermal power law component. This could be the 
inverse-Compton scattered emission of the SS\,433 accretion disk funnel, which is 
collimated in a cone before the scattering off relativistic electrons
--- so called emission cone, hypothetical and invisible directly. This model would remove the
synchrotron model problems of the X-ray emission: of acceleration of the emitting electrons to 
extreme Lorenz factors $\gamma > 10^9$ in the mildly relativistic SS\,433 jets and morphological
difference of W\,50 in the synchrotron radio and X-ray bands. Our study of the comptonization
model showed up that in the case of the minimal factor $\gamma_{\rm min}$ of the order 1
the energy of the relativistic particles, which upscatter the cone photons, is close to or exceeds 
total energy of the nebula --- the case of inapplicability of the model.
In the case of  $\gamma_{\rm min} > 10$ the cone emission is comptonized beyond the 1 -- 10\,keV
band, consequently the existing observational data are non-sensitive to the emission cone.
\end{abstract}

\vspace{0.7cm}

Morphology of X-ray emission of the radio supernova remnant W\,50 in its genesis is connected
with the jets of the microquasar SS\,433. It is seen on the images taken with XMM-Newton 
( \cite[2007]{Bri07}) the hard X-ray emission in the eastern area of W\,50 is localized closer
to SS\,433 and jets axis than the soft X-ray emission. Numerous studies discriminate thermal 
and non-thermal components of the X-ray emission. The thermal component is of non-equilibrium 
ionization, with a temperature $k_{\rm B}T = 0.2-0.3$\,keV. The non-thermal component in 
the brightest knot, lying at the eastern jet axis, of a size $3'\times 7'$, has following characteristics: 
a photon index $\Gamma = 2.41 \pm 0.09$ in the  0.5 -- 8\,keV band, a flux 
$F \approx 5\cdot 10^{-12}$\,erg/cm$^2$\,s and a luminosity $\geq 2\cdot 10^{34}$\,erg/s
in the  0.5 -- 10\,keV band (\cite[1983]{Wat83}; \cite[1994]{Yam94}; \cite[2007]{Bri07}).
In the  0.2 -- 2\,keV band fluxes of both components are comparable.

Synchrotron model of the non-thermal component requires a magnetic field $B \sim 4\,\mu$G
and corresponding factors $\gamma > 10^9$. Such big gammas seem hardly to be in
the mildly relativistic jets of SS\,433. Besides there is not conformity between radio
and X-ray detailed structures of W\,50, therefore rather different emission mechanisms
are responsible for them.

The inherent property of the supercritical accretion, which is acting in SS\,433, is
powerful outflow of gas, of a kinetic luminosity $L_{\rm k} \sim 10^{39}$\,erg/s, 
and concomitant radiation along the channel perpendicular to the 
accretion disk. The radiation, emitting from the disk center and being beamed in a cone,
should be more intensive and hard, than the visible radiation for the observer placed outside
the channel. According to various estimates the emission cone in SS\,433 has an
opening $\Omega_{\rm c} \sim 1$\,sr, an intensity $I_{\rm c} \sim 10^{40}$\,erg/s\,sr,
a soft X-ray thermal spectrum with a temperature $k_{\rm B}T \sim 0.1$\,keV
( \cite[1993]{Pan93}; \cite[1993]{Ara93}; \cite[2006]{Beg06}).

The power law component of the W\,50 X-ray emission associated with the eastern jet 
may be created by inverse-Compton scattering of the cone photons by relativistic electrons 
accelerated during deceleration and interaction of the jet with ambient medium and 
acquiring a power law energy distribution. In approximation of the Thomson scattering 
by the electrons, which have the $\delta$-distribution in energy and the total energy 
\beq
W_{\rm e} = N_{\rm e} m_{\rm e} c^2,
\label{ener}
\eeq 
the luminosity of scattered radiation is 
\beq
L = \frac{4 \gamma^2 \sigma_{\rm T} N_{\rm e} I_{\rm c}}{3 R_{\rm X}^2},
\label{lum}
\eeq
where $\sigma_{\rm T}$ is the Thomson cross-section, $N_{\rm e}$ the total number of
the electrons, $R_{\rm X}$ the distance from the source of seed photons to the scattering region.
The energy range of the inverse-Compton emission will be $\gamma^2$ times the 
seed photons energy range. Hence spectrum of the SS\,433 emission cone should be placed 
in the soft X-ray band and the factor $\gamma_{\rm min}$ should be no more than some.

\begin{figure}[ht]
\centerline{\hbox{\includegraphics[width=14cm]{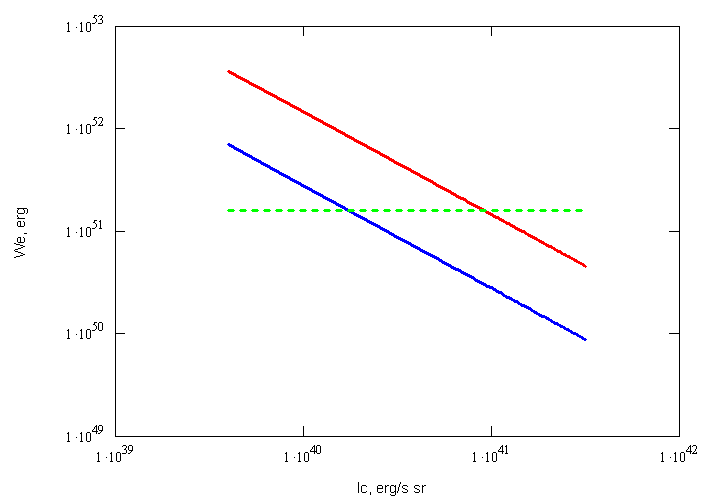}}}
\caption{
The dependency of the total electron energy $W_{\rm e}$ in the bright X-ray knot of the nebula W\,50 
associated with the eastern jet of SS\,433, which is sufficient to pump by  
comptonization an observed flux $F_{\rm 0.5 - 10\,keV} = 5\cdot 10^{-12}$\,erg/cm$^2$\,s 
with a photon index $\Gamma = 2.4$, on the intensity of the SS\,433 cone emission, 
having a temperature $k_{\rm B}T = 0.1$\,keV. The up (red) and bottom (blue) lines
correspond to minimal Lorentz factors $\gamma_{\rm min} = 2$ and 5. The horizontal dashed line
marks the energy released by the jet for a jets lifetime of $5\cdot 10^4$\,yr --- this impose 
the upper limit on value $W_{\rm e}$.
}
\label{IcW}
\end{figure}

The equations (\ref{ener}, \ref{lum}) give a sense of the dependency of the total electron energy on 
the observed flux $F$, hypothetical intensity $I_{\rm c}$ and energy scale of the electrons $\gamma$:
\beq
W_{\rm e} \propto \frac{F}{\gamma I_{\rm c}}.
\label{ener2}
\eeq 
Exact calculation of $W_{\rm e}$ for the case of the power law energy distribution of the relativistic
electrons ${\rm d}N(\gamma) = N_0 \gamma^{-p} {\rm d}\gamma$, where
$p = 2\Gamma - 1$, is shown in Fig.~\ref{IcW} as a function of the cone intensity $I_{\rm c}$. 
Bounds of the distribution range $\gamma_{\rm min} - \gamma_{\rm max}$ are chosen 
so that they will not affect the spectrum slope in the observed range. There are shown the
curves for the bright X-ray knot of W\,50 centered on the eastern jet, for factors  
$\gamma_{\rm min} = 2$ and 5. The latter is  the extreme value, which has 
not yet significantly distorting the observed X-ray spectrum. The factor $\gamma_{\rm max}$ 
almost does not affect the result.

%
%

An energy $L_{\rm k} t_{\rm SS\,433} \sim 10^{39}\,{\rm erg/s} \times 5\cdot 10^4\,{\rm yr}$,
stored in the W\,50 nebula for lifetime of engine of the jets, majorizes possible values of the total
energy of the relativistic electrons responsible for the observed power law X-ray emission
of W\,50 via comptonization mechanism. Really the volume of the emission knot under 
consideration is only $10^{-3}$ part of the W\,50, therefore the above restriction should be
more severe. It can be seen from Fig.~\ref{IcW} the electron energy pool of W\,50 is 
insufficient to energize the power law emission via comptonization under a reasonable cone 
intensity $I_{\rm c} = 10^{40}$\,erg/s\,sr and supposed range of the parameters of SS\,433.

\end{document}